# Implementation of phase gates using single photons

V. Hizhnyakov

Institute of Physics, University of Tartu, W. Ostwald Street 1, 50411 Tartu, Estonia

E-mail : hizh@ut.ee

**Abstract.** Quantum computing algorithms using the quantum Fourier transform require repeated use of a phase shift gate. In the case of qubits using optical photons for operation, this gate can be implemented using single-photon beams focused close to the diffraction limit.

## 1. INTRODUCTION

Single-qubit gates that rotate the state vector and two-qubit CNOT gate represent a universal set of operations that allows for any transformation of the states of quantum computer qubits, so their use is sufficient for quantum computing [1-4]. Hadamard, NOT, and phase shift gates are commonly used as basic single-qubit quantum logic gates. If free or impurity atoms (ions) with different electronic states are used as qubits, then these gates can be implemented using electromagnetic pulses with a certain phase and field strength, the so-called pi type pulses [4-7]. All the main gates, with the exception of the phase gate, lead to a change in the population of qubit levels. In contrast, the phase gate, which is repeatedly used for the quantum Fourier transform (QFT) and the Shore factorization algorithm [1-4], does not change the population of levels. Here we show that maintaining the occupancy of cube levels allows the use of single photons or a small number of photons to realize phase gates.

The mechanism of phase shift of qubit states under consideration resulting in phase gate has a purely quantum origin: it is based on the fact that the interaction of the electromagnetic field with a two-level system leads to the formation of dressed states of the atom and photons [5,7,8]. The energy of dressed states differ from the energy of the initial and final (undressed) states of atom and photons. This difference causes a change of phase of the states of a two-level system after photons pathing it. This change is different for different states of qubit. Here we show that the resulting phase difference can be large enough to realize required phase shifts, even if pulses of single photons or a small number of photons are used, assuming that they are focused on or close to the diffraction limit. Note that a mechanism similar to the phase shift described here causes a finite change in the polarization of elliptically polarized photons by one oriented atom with a rotation angle $\lesssim 27^o$ [9,10].

The implementation of phase gates using single photons may be much more practical than using pi type pulse sequences for this purpose. This may be especially true for the Shor algorithm that uses the quantum Fourier transform for a large number $N = 2^n$ of qubit states ($n$ is the number of qubits). Here we take into account that the implementation of QFT may require the management of a very large number of phases. This is difficult to achieve using pi type pulses, but at the same timeit is much easier to achieve with single photons. In this communication, we will discuss the implementation of phase gates using single photons for impurity centers in crystals with different electronic states acting as qubits. Our consideration primarily concerns optical qubits, considered in [11-14], based on impurity centers of trivalent rare earth ions. However, this also applies to quantum computers with trapped ions and other quantum computers with qubits that use optical photons to operate.

## 2. PHASE GATE WITH PI PULSE

Among the basic single qubit gates most important are the Hadamard $H$, $NOT \equiv X$, $Y$ and phase (shift) $Z(\theta)$ gates. In the basis of qubit states $|0\rangle$ and $|1\rangle$, these gates are described by the following unitary matrices (operators):

$$H = \frac{1}{\sqrt{2}}\begin{pmatrix} 1 & 1 \\ 1 & -1 \end{pmatrix}, \quad X = \begin{pmatrix} 0 & 1 \\ 1 & 0 \end{pmatrix}, \quad Y = \begin{pmatrix} 0 & -i \\ i & 0 \end{pmatrix}, \quad Z(\theta) = \begin{pmatrix} 1 & 0 \\ 0 & e^{i\theta} \end{pmatrix}. \tag{1}$$

First three of these operators are not diagonal, and their application leads to a change of the population of the qubit levels and qubit energy; the last oprator, corresponding to the rotation of the state vector in the Bloch sphere around the $z$ axis conserves the populations. The use of this operator is the basis of QFT: a large number of dyadic rational phase gates $R_k = Z(\theta = 2\pi/2^k)$, $k = 0,1,2,...N-1$ are implemented in QFT.

First, we will consider the implementation of a phase gate $Z(\theta)$ for a two-level system with qubit states $|0\rangle$ and $|1\rangle$ when excited by a strong quasi-monochromatic light beam. The Hamiltonian of this system interacting with a classical electromagnetic field has the following form [4,5]

$$H_A = \hbar\left(\omega_1(I-\sigma_z)/2 + \kappa E(t)\sigma_x\right), \tag{2}$$

where $\sigma_x \equiv X$ and $\sigma_z \equiv Z(0)$ are the Pauli matrices, $I$ is the unite matrix, $\omega_1$ is the frequency difference of the qubit states, $E(t) = \mathcal{E}_t \cos(\omega t + \varphi)$ is the electric field strength of the exciting light pulse at time $t$, $\mathcal{E}_t$ is the amplitude of the electric field of the pulse, which is slowly time-dependent, $\kappa = 2e\langle 0|r|1\rangle/\hbar$ is the interaction constant of the elctric field of the pulse with the qubit, $\langle 0|r|1\rangle$ is the matrix element of the dipole transition in the qubit, $\omega = \omega_1 + \Delta$ is the average frequency and $\varphi$ is the initial phase of the light pulse, In the case of a quasi-monochromatic pulse of light, the rate of amplitude $\mathcal{E}_t$ change over time is small compared to the average frequency $\omega$: $|\dot{\mathcal{E}}_t| \sim \delta \ll \omega$.

We take the time dependence of the wave function of a two-level system as

$$|\psi(t)\rangle = \begin{pmatrix} C_0(t) \\ C_1(t)e^{-i\omega_1 t} \end{pmatrix}, \tag{3}$$

where $C_0(t)$ and $C_1(t)$ are the time-dependent amplitudes of the states $|0\rangle$ and $|1\rangle$ of the qubit, respectively. From the Schredinger equation $i\hbar|\dot\psi\rangle = H|\psi\rangle$ it follows that

$$\dot{C}_0(t) = -i\kappa E(t)C_1(t)e^{-i\omega_1 t}, \quad \dot{C}_1(t) = -i\kappa E(t)C_0(t)e^{i\omega_1 t}. \tag{4}$$

*Resonant excitation*

Consider first the excitation of a two-level system by a laser pulse with an average frequency $\omega$ close to the qubit frequency $\omega_1$. In this case, we can neglect the rapidly fluctuating terms $\propto e^{\pm i(\omega+\omega_1)t}$. As a result, we obtain the following equations [4,7]

$$\dot{C}_0(t) = -i\kappa\mathcal{E}_t C_1(t)e^{i\Delta t + i\varphi}, \quad \dot{C}_1(t) = -i\mathcal{E}_t C_0(t)e^{-i\Delta t - i\varphi} \tag{5}$$

In the case of the strict resonance $\Delta \equiv \omega - \omega_1 = 0$, we get

$$C_0(t) = C(0)\cos\Theta_t - iC_1(0)\sin\Theta_t,$$
$$C_1(t) = \left(-iC_0(0)\sin\Theta_t + C_1(0)\cos\Theta_t\right)e^{-i\varphi}. \tag{6}$$

where $C_0(0)$ and $C_1(0)$ are the initial (before the light pulse) values of the population amplitudes of the qubit states $|0\rangle$ and $|1\rangle$, $\Theta_t = \kappa \int_{-\infty}^{t} \mathcal{E}_{t'} dt'$ is the rotation angle of the state under consideration in the Hilbert space.

The state of a two-level system after the pulse has passed is described by the rotation angle $\Theta = \kappa \int_{-\infty}^{\infty} \mathcal{E}_t dt$. Using resonance light pulses, the NOT gate is implemented in case of $\Theta = \pi$ and arbitrary $\varphi$; the Hadamard gate $H = (\sigma_x + \sigma_z)/\sqrt{2}$ is implemented if $\Theta = \pi/4$ and $\varphi = -\pi/2$. The implementation of the phase gate $Z(\theta)$ is also possible if a light pulse with $\Theta = \pi$ and $\theta = -\varphi$ is used. It should be noted that the control of the initial phases $\varphi$ of the laser pulses requires the use of interferometric devices.

*Excitation near resonance*

Let us now consider the case of excitation near the resonance when $|\Delta| > |\kappa \mathcal{E}_t|$. Using reduced amplitudes $\tilde{C}_{0,1} = C_{0,1} e^{\pm(i\Delta t + \varphi)/2}$, equations (5) can be represented as

$$\dot{\tilde{C}}_0 \approx -i\Delta \tilde{C}_0/2 - i\kappa\mathcal{E}_t \tilde{C}_1, \quad \dot{\tilde{C}}_1 \approx i\Delta \tilde{C}_1/2 - i\kappa\mathcal{E}_t \tilde{C}_0. \tag{7}$$

In the case $|\Delta| > |\kappa \mathcal{E}_t|$, approximate solutions of these equations read $\tilde{C}_{0,1} \approx e^{\pm it(\Delta/2(1+2\kappa^2 e_t^2/\Delta^2))}$, which gives

$$C_0(t) \approx C(0)\cos\tilde{\Theta}_t - iC_1(0)\sin\tilde{\Theta}_t,$$
$$C_1(t) \approx \left(-iC_0(0)\sin\tilde{\Theta}_t + C_1(0)\cos\tilde{\Theta}_t\right)e^{-i\varphi}, \tag{8}$$

where $\tilde{\Theta}_t = (\kappa^2/\Delta)\int_{-\infty}^{t} \mathcal{E}_{t'}^2 dt'$ is the rotation angle of the states under consideration in the Hilbert space. Note that this angle is smaller than that in the case of resonant excitation. Therefore, to implement NOT, Hadamard, phase, and ohter quantum gates, using near-resonant excitation, it is necessary to apply stronger pulses than in the case of strictly resonant excitation.

Although the use of pi type pulses with initial phase control allows the implementation of basic quantum gates, including the phase shift gate, the required phase control of the pulses is challenging. Indeed, to implement QFT in the Shore large number factorization algorithm, it is necessary to use a large number $n \gg 1$ of qubits. In this case the $QFT$ operator contains very large number $N^2 = 2^{2n}$ of elements, each of which has its own phase. In order to control the phase of each of these elements using the pi type pulses, it is necessary to supply at least $2 \cdot N^2$ such pulses. To do this, it is necessary to control the initial phases of all these pulses with high accuracy. To this end, various interferometric devices must be used to control a large number of initial phases of the laser pulse. All this makes the implementation of QFT with pi type pulses a rather difficult task.

## 3. IMPLEMENTATION OF A PHASE GATE WITH SINGLE PHOTONS

Here, the possibility of implementing phase gates using irradiation of qubits with weak light beams, including single-photon ones, is considered. We will show that such beams allow us to obtain phase corrections of qubit states, which are necessary for phase control in the implementation of QFT and other gates. This is possible when focusing photon beams near the diffraction limit. We take into account that as a result of the interaction of a two-level system with photons focused on it, entangled, dressed states of this system and photons are formed [5,7,8]. These states cause a phase change, which makes it possible to achieve the required phase difference between the states of the qubits.

Dressed states $|\psi_0\rangle_{N_0}$ and $|\psi_1\rangle_{N_0}$ are characterized by the number of excitations $N_0$, including the number of photons and the number excitatations of the two level system, and they can be presented as the mixture of the state $|0\rangle|N_0\rangle$ with $N_0$ photons and zeroth state of the two level system and the state $|1\rangle|N_0-1\rangle$ with $N_0-1$ photons and the first state of the two level system. We emphasize that states with similar energy are mixed.

We consider the case when a two-level system is excited by light pulse consisting of $N_0$ quasi-monochromatic photons. In this case, the average wave functions of dressed states with $N_0$ excitations have the following form [7.8]:

$$|\Psi_0\rangle_{N_0} = -\sqrt{\frac{\Omega_{N_0} - \Delta}{2\Omega_N}}|1\rangle|N_0-1\rangle + \sqrt{\frac{\Omega_{N_0} + \Delta}{2\Omega_{N_0}}}|0\rangle|N_0\rangle,$$

$$|\Psi_1\rangle_{N_0} = \sqrt{\frac{\Omega_{N_0} + \Delta}{2\Omega_{N_0}}}|1\rangle|N_0-1\rangle + \sqrt{\frac{\Omega_{N_0} - \Delta}{2\Omega_{N_0}}}|0\rangle|N_0\rangle,$$

(9)

where

$$\Omega_{N_0} = \sqrt{\Delta^2/4 + \kappa^2 \varepsilon_{N_0}^2} \tag{10}$$

is the Rabi frequency,

$$\varepsilon_{N_0} = |\langle N_0 - 1|\hat{E}|N_0\rangle| \cong \sqrt{\hbar \omega N_0 Z/t_0 d^2}/n \tag{11}$$

is the value of the matrix element of the electric field operator $\hat{E}$ for average dressed states, $d^2$ is the cross-section of the light beam, $t_0 = \delta^{-1}$ is the flight time of the light beam through the two-level system, $Z = 376.7$ ohm is the impedance of free space, $n$ is refractive index. Note that states (9) are entangled. The average change in the frequency of the dressed states $|\psi_{0,1}\rangle_{N_0}$ caused by the interaction of a two-level system with a wave paket of $N_0$ photons equals

$$\delta\omega_{0,1}^{(N_0)} \cong \pm\left(\Omega_{N_0} - \Delta/2\right). \tag{12}$$

The sign of frequency change corresponds to the $\Delta$ sign.

We consider the case when a two-level system is excited by a pulse of quasimonochromatic photons with a spectral width significantly smaller than the average Rabi frequency. In this case $t_0 \gg \Omega_{N_0}^{-1}$, and the adiabatic regime of the switching on and off of the interaction of the two-level system with photons takes place. Therefore the initial state $|0\rangle$ of the two-level system is adiabatically transformed to the state $|\Psi_0\rangle_{N_0}$ and finally back into the state $|0\rangle$, but with a phase shift. Similarly, the initial state $|1\rangle$ of the two-level system is adiabatically transformed to the state $|\Psi_1\rangle_{N_0+1}$ and finally back into the $|1\rangle$ state, but with a phase shift. The change of phase of the initial state $|0\rangle$ of a two-level system caused by its interaction with the quasi-monochromatic $N_0$-photon beam is equal to $\delta\omega_{0;N_0} t_0 \cong (\Omega_{N_0} - \Delta/2) t_0 \operatorname{sgn}\Delta$. Similarly, the change of phase of the state $|1\rangle$ of a two-level system caused by its interaction with the quasi-monochromatic $N_0$-photon beam is equal to $-(\Omega_{N_0+1} - \Delta/2) t_0 \operatorname{sgn}\Delta$. Note that the signs of the phase changes of the states $|0\rangle$ and $|1\rangle$ are different. The phase difference in the case of $\Omega_{N_0} - |\Delta/2| \gtrsim |\Delta/2|$ is

$$\varphi_{N_0} \sim |e\langle 0|r|1\rangle/\hbar d|\left(\sqrt{N_0} + \sqrt{N_0+1}\right)\sqrt{\hbar\omega Z t_0}\ \operatorname{sgn}\Delta \tag{13}$$

Now we take into account that the light beam can be can be maximally focused within the diffraction limit. In this case $d = \lambda/2n$, where $\lambda$ is the wave length of light, $n$ is the refraction index. Therefore the maximum possible value of the phase difference of qubit states caused by a sigle photon is

$$|\varphi_1^{max}| \sim 2|e\langle 0|r|1\rangle|\varepsilon_1^{max} t_0/\hbar = 4.8|e\langle 0|r|1\rangle/\lambda|\sqrt{\omega Z t_0/\hbar}, \tag{14}$$

where $\varepsilon_1^{max} = 2.4\sqrt{\hbar\omega Z/(t_0\lambda^2)}$ is the maximum value of the electric field produced by a single-photon beam of the duration $t_0$ when it is focused at the diffraction limit $(\lambda/2n)^2$.

## 4. DISCUSSION

In [10-14], mixed and heterovalent crystals with impurity centers of rare earth element ions were proposed for ultrafast quantum computers with different 4f states as qubits of optical frequencies. It was found that among these centers there are those in which there are simultaneously states in which the centers interact weakly, and states in which the centers interact strongly, provided that the concentration of centers is large enough. The first states are suitable for qubit states, while the second states can be used as auxiliary states for implementing CNOT and other conditional gates. The large inhomogeneous broadening of the spectra of electronic transitions in mixed crystals makes it possible to distinguish a large number of qubits when using ultrashort light pulses to control the qubits. The use of ultrafast quantum gates can reduce the impact of coherence loss, which can allow quantum computing to be performed without using very low temperatures.

As an example of the implementation of phase gates using single photons for qubits with ions of rare earth elements, we consider here 4f levels $^3H_6$ and $^1G_4$ of $Tm^{3+}$ ion in $CaF_2$ crystal as a two level system (these states were considered in [12] as possible states of qubits). Corresponding electronic transition has energy $\hbar\omega_1 = 2.63$ eV ($\lambda = 472.3$ nm) and radiative rate $\gamma = 0.91\cdot 10^3$ s$^{-1}$ [15]. In a free ion, this transition is dipole-forbidden; in a crystal, it is partially dipole-allowed due to the low symmetry of the $CaF_2$:$Tm^{3+}$ center. The magnitude of the dipole matrix element of this transition induced by the low symmetry crystal field can be estimated as follows:

$$|\langle 0|r|1\rangle| = \sqrt{3\gamma_0/4\alpha c k^3} \approx 6.0\cdot 10^{-10}\ \text{cm} \tag{15}$$

(here $\alpha = e^2/\hbar c$ is the fine-structure constant). A single photon beam of the duration $t_0 = 10^{-7}$ s focused on the diffraction limit can produce the electric field

$$\varepsilon_1^{max} \sim 2.4\sqrt{\hbar\omega Z/(t_0\lambda^2)} \approx 28.9 \text{ V/cm}. \qquad (16)$$

Note also that in the case under consideration $\Omega_{N_0} \sim (\sqrt{N} + \sqrt{N+1})\cdot 10^7 \text{ s}^{-1}$. This exceeds $t_0^{-1} = 10^7 \text{ s}^{-1}$. This means that the requirement of adiabatically switching on and off the interaction of photon pulse with a two-level system is fulfilled. The maximum phase difference of qubit states, which can be produced by such a photon in the case of the electronic transition under considseration is

$$|\varphi_1^{max}| \sim 4.8|e\langle 0|r|1\rangle|\varepsilon_1^{max}t_0/\hbar \simeq 10.6 \text{ radian} \qquad (17)$$

This is the limit value of the phase difference, which is difficult to achieve due to the difficulty of achieving maximum focusing with a $(\lambda/2n)^2$ cross section. However, to implement QFT, it is enough to focus ~100 photons at $(3\mu)^2$, which is easier to achieve. Even worse focusing requires the use of light pulses with a large number of photons. In addition, shorter pulses require better focusing and/or more photons. In particular, increasing (decreasing) the number of photons by several times often makes it possible to reduce (increase) the pulse duration by ten times.

Here the qubit transitiion $|4f\rangle \to |4f'\rangle$ between 4f states was used to implement the phase gate. However, the electronic tgransirtiions $|4f\rangle \to |j\rangle$ or $|4f'\rangle \leftrightarrow |j'\rangle$ between 4f qubit levels and other ion levels can also be used to implement the phase gate by changing the phase of only one of the qubit states. In particular, the $5d$ states can also be used as $|j\rangle$ and $|j'\rangle$ states. Electronic transitions $|4f\rangle \to |5d\rangle$ are dipole allowed; the dipole matrix elements $|\langle 4f|r|4d\rangle|$ of these transitions are three or more orders of magnitude larger than the dipole matrix elements $|\langle 4f|r|4f_1\rangle|$ of impurity centers. Therefore, the phase change $|\varphi_1^{max}|$ for $|4f\rangle \to |5d\rangle$ transition should also be three or more orders of magnitude greater than for $|4f\rangle \to |4f'\rangle$ transitions. As a result, single photon beams with a frequency close to resonant with a transition $|4f\rangle \to |5d\rangle$ and fairly moderate focusing can be used to implement phase gates.

It should be noted that NV, SiV and analogous centers in diamond are also considered as a promising sytems for quantum computers [16,17]. These centers have dipole allowed electronic transitions in the visible range with narrow spectral lines, which are suitable for laser control of qubits at spin sublevels. These transitions are also well suited for implementation of phase gates. For example, a transition with a wavelength 737 nm in SiV centers has a life-time limited spectral width 100 MHz. The dipole matrix element of this transition equals $6\cdot 10^{-7}$ cm, which is three orders of magnitude larger than in rare-earth ions. Correspondingly, $|\varphi_1^{max}|$ is also approximately three orders of magnitude larger. Therefore, single photon beams with a wavelength of 737 nm can be used to implement phase gates in SiV centers.

Finally, we note that trapped-ion quantum computers also use dipole allowed optical transitions for operation of qubits. For example Ca$^+$ ion-based trapped-ion quantum computer uses for operation $4^2S_{1/2} \to 4^2P_{1/2}$, $4^2P_{3/2}$ transitions (396.847 nm, 7.7 ns and 393.366 nm, 7.4 ns) [18] with three orders of magnitude larger dipole matrix elements than in rare-earth ions. These transitions also make it possible to implement phase gates using single photons with fairly moderate focusing.